\documentclass[aip,nofootinbib,12pt]{revtex4-1}
\usepackage{amssymb,amsmath,amsthm}
\usepackage{textcomp}

\usepackage{color}
\usepackage{latexsym}
\usepackage[english]{babel} 
\usepackage[latin1]{inputenc}  
\usepackage[T1]{fontenc}   
\usepackage[all]{xy}
\usepackage{amsfonts}
\usepackage{graphics}
\usepackage{epsfig}
\usepackage{graphicx}
\usepackage[colorlinks=true,linkcolor=blue]{hyperref}%

\def\RR{\mathbb{R}}
\def\CC{\mathbb{C}}

\def\C{\mathbb{C}}

\def\QQ{\mathbb{Q}}
\def\PP{\mathbb{P}}
\def\SS{\mathbb{S}}
\def\OO{\mathbb{O}}

\def\AA{\mathbb{A}}

\def\HH{\mathbb{H}}
\def\BB{\mathbb{B}}

\usepackage{tikz}
\usetikzlibrary{automata}
\usetikzlibrary{shadows}
\usetikzlibrary{arrows}
\usetikzlibrary{shapes}
\usetikzlibrary{fit}
\usetikzlibrary{matrix}
\newcommand{\incl}{\ar@{^{}-}}
\newcommand{\inclu}{\ar@{^{}.}}

\def\e{\epsilon}

\newtheorem{prop}{Proposition}[section]

\newtheorem{def }{Definition  }[section]
\newtheorem{theorem}{Theorem}
\newtheorem{ex}{Example     }[section]
\newtheorem{rem}{Remark    }[section]

\begin{document}

 \title{Classification of multipartite systems featuring only $|W\rangle$ and $|GHZ\rangle$ genuine entangled states}
\author{Fr\'ed\'eric Holweck\footnote{frederic.holweck@utbm.fr, 
Laboratoire IRTES-M3M,
Universit\'e de Technologie de Belfort-Montb\'eliard, 90010 Belfort Cedex, France}, 
P\'eter L\'evay\footnote{levay@phy.bme.hu, Department of Theoretical Physics, Institute of Physics, Budapest University of Technology and Economics
, H-1521 Budapest, Hungary}\footnote{Hungarian Academy of Sciences}}

 \begin{abstract}
In this paper we present several multipartite quantum systems featuring the same type  of genuine 
(tripartite) entanglement. Based on a geometric interpretation of the so-called $|W\rangle$ and $|GHZ\rangle$ states 
we show that the classification of all multipartite systems featuring those and only 
those two classes 
of genuine entanglement can be deduced from earlier work of algebraic geometers. This classification corresponds in fact to  classification 
 of fundamental subadjoint varieties and establish a connection between those systems, well known in Quantum Information Theory and fundamental simple Lie algebras. 
 \end{abstract}

\maketitle

\section{Introduction}

Entanglement is a key resource of quantum information. It corresponds to a form of 
correlation between subsystems  of 
a given composite system which is stronger than any correlation arising from  classical communication\cite{bell}.
Since the advent of Quantum Information a large amount of experimental and theoretical evidence demonstrates that quantum protocols featuring this phenomenon exist that can overperform their classical 
counterparts. 
Beyond this  
entanglement is of basic importance for obtaining new communication protocols such as quantum teleportation,  quantum superdense codding or quantum cryptography. It is also acknowledged that 
quantum entanglement plays a central role in quantum algorithms and quantum computation\cite{nielsen,KSV}.

Quantum entanglement is a consequence of the superposition principle. 
Let us illustrate this for a bipartite system.
Given two copies of two-state systems (qubits) represented by the vectors  $|\psi\rangle_A\in \mathcal{H}_A$ and $|\psi\rangle_B\in \mathcal{H}_B$ with $\mathcal{H}_A\simeq \mathcal{H}_B\simeq \CC^2$, we define the composite system of two qubits as the one represented by the tensor product $\mathcal{H}=\mathcal{H_A}\otimes\mathcal{H}_B$. 
Then a canonical basis of $\mathcal{H}$ is 
$\{|00\rangle, |01\rangle, |10\rangle, |11\rangle\}$ and the superposition principle 
tells us that one possible state for the composite system 
is
\begin{equation}
 |\psi\rangle_{AB}=\dfrac{1}{\sqrt{2}}(|00\rangle+|11\rangle).
\end{equation}
\noindent
However, $\vert\psi\rangle_{AB}$ cannot be created 
from an initial state of type $\vert\varphi\rangle_A\otimes \vert\chi\rangle_B$ by applying only local operations (i.e. operations 
acting on $|\varphi\rangle_A$ and $|\chi\rangle_B$ separately). Such states are called entangled. Since entangled states cannot be generated 
locally they correspond to  a global resource shared 
by  the actors of the protocol. On the other hand a state is said to be separable if it can be generated locally from a state of the form $\vert\varphi\rangle_A\otimes \vert\chi\rangle_B$.

The multipartite generalization of our example provides the basic resource for Quantum Information. 
However, as a resource entanglement needs to be classified. 
One possible classification scheme is obtained by finding equivalence classes of entangled states under Stochastic Local Operations and Classical Communications (SLOCC). SLOCC transformations are the mathematical representatives of certain physical manipulations allowed to be performed on our composite system. These manipulations are consisting of local reversible operations on each component of the multipartite system
assisted by classical communication (i.e. the local operations may be coordinated).
The word stochastic refers to the possibility of converting a particular state of the system to another one  and vice versa with some (generally different) probability of success.
The mathematical representative of such SLOCC transformations turns out to be a group acting on the Hilbert space of the composite system. The precise form of this group will depend on the observables characterizing this system.

The classification of entanglement classes of multipartite systems under SLOCC has 
been investigated in the last 10 years by many authors\cite{Dur,brody2,BDDER,My,My2,My3,VDMV,HLT,HLT2,LV,VL,BDD,DF,DF2,SL}. 
Interestingly under SLOCC some\cite{brody2,Dur,LV,VL,SL,DF,BDDER}
of these entangled systems is featuring two genuine types of entanglement.
The aim of this paper for these systems is to provide 
a unified approach based on recent results of algebraic geometry.

The paper is organized as follow. In Section \ref{3qubit} we introduce a geometric interpretation\cite{HLT,HLT2} of the entangled classes $|W\rangle$ and $|GHZ\rangle$ which 
correspond to the two genuine entangled classes in the D\"ur, Vidal and Cirac\cite{Dur}'s classification
of entanglement classes of three qubits. Thanks to this geometric interpretation we can use, in Section \ref{tripartite}, classical results of invariant theory and 
algebraic geometry to classify all Hilbert spaces and quantum systems which will feature
those two and exactly those two types of genuine entangled classes. In this process we recover different quantum systems investigated in the quantum information literature 
and three new cases.  The corresponding Hilbert spaces   
have a  similar SLOCC orbit structure (except for the case of three bosonic qubits see Remark \ref{bosons}). We also classify quantum systems with two and exactly two entanglement classes (not necessarly of type $|W\rangle$ and $|GHZ\rangle$). 
The connexion of those quantum systems 
with classification results from algebraic geometry allows to give a uniform description of those quantum systems and establish a link between those systems and the simple Lie algebras.
In particular we collect some geometrical information about such system in  Appendix \ref{app}. 

{\bf Notations.}
In the text $V$ (resp. $\mathcal{H}$) will denote a vector (resp. Hilbert) space over the field of complex numbers $\CC$, and $\PP(V)$ (resp. $\PP(\mathcal{H})$) will denote the corresponding projective space. A vector $v\in V$ will be projectivized to a point $[v]\in \PP(V)$.
A projective algebraic variety $X\subset \PP(V)$ will be defined as the zero locus of a collection of homogeneous polynomials. A point $[x]\in X$ will be said to be smooth if and only if the partial derivatives of the defining 
equations do not simultaneously vanish at $[x]$. If $[x]\in X$ is smooth, one can define $\tilde{T}_x X\subset \PP(V)$ the embedded tangent space of $X$ at $[x]$.

In this article we only consider pure quantum systems, i.e. a state $|\psi\rangle$ of such systems will always be considered as a (normalized) vector of $\mathcal{H}$.

\section{The three qubit classification revisited}\label{3qubit}
Starting from the paper of D\"ur, Vidal and Cirac \cite{Dur} three-qubit entanglement has given rise to a number of interesting applications 
\cite{CKW,KL,BDDER2,BDL,Levay1}. 
Let us denote by $\mathcal{H}_A$, $\mathcal{H}_B$ and $\mathcal{H}_C$ the three Hilbert spaces isomorphic to $\CC^2$ corresponding to
qubits $A$, $B$ and $C$, then the Hilbert space of the composite system is
$\mathcal{H}=\mathcal{H}_A\otimes \mathcal{H}_B\otimes \mathcal{H}_C$.
In this section for simplicity let us adopt the notation $\vert\psi\rangle_A\equiv \psi_A$. If we forget about scalar normalization the relevant SLOCC group turns out to be
$GL_2(\CC)\times GL_2(\CC)\times GL_2(\CC)$ and
the
result established in Ref\cite{Dur} states that under SLOCC action 
three qubits can be organized into six orbits i.e. SLOCC entanglement classes (Table~\ref{table_3q}). 
\begin{table}[!h]
 \begin{center}
  \begin{tabular}{c|c}
Name & Normal form \\
\hline
Separable & $|000\rangle$\\
\hline
Biseparable &$\dfrac{1}{\sqrt{2}}(|000\rangle+|011\rangle)$\\

Biseparable & $\dfrac{1}{\sqrt{2}}(|000\rangle+|101\rangle)$\\

Biseparable & $\dfrac{1}{\sqrt{2}}(|000\rangle+|110\rangle)$\\
\hline
$|W\rangle$ & $\dfrac{1}{\sqrt{3}}(|100\rangle+|010\rangle+|001\rangle)$ \\
\hline
$|GHZ\rangle$ & $\dfrac{1}{\sqrt{2}}(|000\rangle+|111\rangle)$
\end{tabular}
\caption{Three qubit classification}\label{table_3q}
\end{center}
\end{table}
The three qubits  classification features the 
interesting property of having exactly two classes of genuine entanglement, called the $|W\rangle$ and $|GHZ\rangle$ classes.
It should be also clear that, for instance, for 
the bipartite state 
$|\psi\rangle=\dfrac{1}{\sqrt{2}}(|000\rangle+|101\rangle)$, particles $A$ and $C$ are entangled while $B$ is not.
Note that from the projective point of view multiplication by a nonzero scalar does not change the nature of the state and thus we can instead consider 
the $SL_2(\CC)\times SL_2(\CC)\times SL_2(\CC)$ orbits of $\PP(\mathcal{H})$. It turns out that this classification was 
known, from mathematical perspective, since the work
of Le Paige\cite{LePai}. 
Motivated by this example  we can adress the basic question of our paper as: which other types of quantum systems 
do have two and 
only two types of genuine non-equivalent entangled states?

Following Ref\cite{HLT} let us rephrase the classification of three qubit entanglement classes by means of algebraic geometry.
In the projectivized Hilbert space $\PP(\mathcal{H})$ we consider the image of the following map:
\begin{equation}
 \begin{array}{cccc}
  \phi :& \PP(\mathcal{H}_A)\times\PP(\mathcal{H}_B)\times\PP(\mathcal{H}_C) & \to& \PP(\mathcal{H})\\
       &   ([\psi_A],[\psi_B],[\psi_C]) & \mapsto & [\psi_A\otimes \psi_B\otimes \psi_C]
 \end{array}
\end{equation}

The map $\phi$
is well-known as the Segre map\cite{Ha,Lan}. Let $X=\phi(\PP(\mathcal{H}_A)\times\PP(\mathcal{H}_B)\times \PP(\mathcal{H}_C))\subset \PP(\mathcal{H})$,
 in what follows we will denote this image simply by $X=\PP^1\times\PP^1\times\PP^1$ because the projectivization of the Hilbert space of each particle clearly corresponds to a projective line ($\PP(\mathcal{H})=\PP(\CC^2)=\PP^1$).
 It can be easily shown that $X$ is a smooth projective algebraic variety\cite{Ha}.
 It is also clear that the variety $X$ is the $G=SL_2(\CC)\times SL_2(\CC)\times SL_2(\CC)$ orbit 
 of any rank one tensor in $\PP(\mathcal{H})$. Indeed given $[\psi_A\otimes \psi_B\otimes\psi_C]$, then for any $[\tilde{\psi}_A\otimes\tilde{\psi}_B\otimes\tilde{\psi}_C]$, there exists 
 $g=(g_1,g_2,g_3)\in G$ such that $[\tilde{\psi}_A\otimes\tilde{\psi}_B\otimes\tilde{\psi}_C]=[(g_1\cdot\psi_A)\otimes (g_2\cdot\psi_B)\otimes (g_3\cdot\psi_C)]$. In terms of 
 quantum entanglement it follows from this description that the variety $X$ represents the set of separable states and can be described 
 as the projectivized orbit of $\psi=|000\rangle$, i.e. $X=G\cdot[|000\rangle]\subset \PP(\mathcal{H})$.
 
 Similarly to what was done in Ref\cite{HLT,HLT2} our goal is now to build from  the algebraic variety of separable states 
 some auxiliary 
 varieties which will encode different type of entanglement classes. Consider $Y^n\subset\PP(V^{n+a+1})$ a projective 
 algebraic variety of dimension $n$ embedded in a projective space of dimension $n+a$, such that $Y$ is smooth and not contained in a hyperplane.
 Then one defines the secant variety\cite{Z2} of $Y$, denoted by $\sigma(Y)$, as the algebraic closure, for the Zariski topology, of the union of secant lines of $Y$ (Eq. (\ref{secant}))
 \begin{equation}\label{secant}
  \sigma(Y)=\overline{\bigcup_{x,y\in Y} \PP^1 _{xy}}
 \end{equation}
Another interesting auxiliary variety  is the tangential variety of $Y$, which is defined as the union of embedded tangent
spaces, $\tilde{T}_y Y$ of $Y$ (Eq. (\ref{tangent}))
\begin{equation}\label{tangent}
 \tau(Y)={\bigcup_{y\in Y} \tilde{T}_y Y}
\end{equation}
One point of importance is the following: If the variety $Y$ is $G$-invariant for the action of a group $G$ (i.e. if $y\in Y$ then $g.y\in Y$ for all $g\in G$) then so are 
the varieties $\tau(Y)$ and $\sigma(Y)$. This property follows from the defintions of the two auxiliary varieties $\sigma(Y)$ and $\tau(Y)$ which 
are built from points of $Y$.

Clearly $\tau(Y)\subset \sigma(Y)$, as tangent lines can be seen as limits of secant lines, and the expected dimensions of those varieties
are $\text{min}\{2n,n+a\}$ for $\tau(Y)$ and $\text{min}\{2n+1,n+a\}$ for $\sigma(Y)$. 
A consequence of the Fulton-Hansen connectedness Theorem\cite{FHa} is the following corollary which will be central to what follows.
\begin{prop}\label{dim}[Corollary 4 of Ref\cite{FHa}]
 One of the following must hold
 \begin{enumerate}
  \item $\text{dim}(\tau(Y))=2n$ and $\text{dim}(\sigma(Y))=2n+1$,
  or
    \item $\tau(Y)=\sigma(Y)$.
 \end{enumerate}
\end{prop}

If we go back to the case where $X=\PP^1\times\PP^1\times\PP^1\subset \PP^7$, a standard calculation\footnote{The dimension of secant variety can be calculated via the Terracini's Lemma. The case we are interested in, is for instance explicitly done in Ref\cite{Lan3} example 5.3.1.5 page 123. Calculations involving 
Terracini's Lemma in the context of QIT can also be found in Ref\cite{HLT,HLT2}} shows that $\text{dim}(\sigma(X))=7$, i.e. the secant
variety is of expected dimension and fills the ambient space. Therefore one automaticaly knows
that $\tau(X)$ exists and is of codimension one in $\PP^7$. Moreover given a general pair of points $(x,y)\in X\times X$ denoted 
by $x=[\psi_A\otimes \psi_B\otimes \psi_C]$ and $y=[\tilde{\psi}_A\otimes \tilde{\psi}_B\otimes \tilde{\psi}_C]$ it is not difficult to see that there exists 
$g=(g_1,g_2,g_3)\in SL_2(\CC)\times SL_2(\CC)\times SL_2(\CC)$ such that $[g.(|000\rangle+|111\rangle)]=[(g_1.|0\rangle)\otimes (g_2.|0\rangle)\otimes (g_3.|0\rangle)+(g_1.|1\rangle)\otimes (g_2.|1\rangle)\otimes (g_3.|1\rangle)=[\psi_A\otimes \psi_B\otimes \psi_C+\tilde{\psi}_A\otimes \tilde{\psi}_B\otimes \tilde{\psi}_C]$.
In other words we have for $G=SL_2(\CC)\times SL_2(\CC)\times SL_2(\CC)$
\begin{equation}\label{secantGHZ}
 \sigma(X)=\sigma(G.[|000\rangle])=\overline{G.[|000\rangle+|111\rangle]}
\end{equation}
Similarly one can provide an orbit description of $\tau(X)$: Let $\gamma(t)=[(\psi_A+t\tilde{\psi}_A)\otimes (\psi_B+t\tilde{\psi}_B)\otimes(\psi_C+ t\tilde{\psi}_C)]$
be a general curve of $X$ passing through $[\psi_A\otimes \psi_B\otimes \psi_C]$ such that $\psi_i$ and $\tilde{\psi}_i$ are not colinear.
Then a straightforward calculation shows that after differentiation we get $\gamma'(0)=[\tilde{\psi}_A\otimes\psi_B\otimes\psi_C+\psi_A\otimes\tilde{\psi}_B\otimes\psi_C+\psi_A\otimes\psi_B\otimes\tilde{\psi}_C]\in \tilde{T}_{[\psi_A\otimes \psi_B\otimes \psi_C]}X$.
Again under the group action $G$ this calculation tells us that 
\begin{equation}\label{tangentW}
 \tau(X)=\tau(G.[|000\rangle])=\overline{G.[|100\rangle+|010\rangle+|001\rangle]}
\end{equation}

Equations (\ref{secantGHZ}) and (\ref{tangentW}) say that the $|GHZ\rangle$ and $|W\rangle$ states form open subsets 
of the secant and tangential varieties of the set of separable states respectively. Moreover, the fact that the secant variety has the expected dimension
implies that those two states are non equivalent.
Considering the biseparabe states the geometric picture can be completed as follows:

\begin{figure}[!h]
\[\xymatrix{&\sigma(X)=  \PP^7&  \\ 
&\tau(\PP^1\times \PP^1 \times \PP^1)\incl[dr]\incl[u] &  \\
    \sigma(\PP^1\times \PP^1)\times \PP^1       \incl[ru] \incl[rd] & \PP^1\times\sigma(\PP^1\times \PP^1)\incl[u] \incl[d]& \sigma(\PP^1\times\underline{\PP^1}\times \PP^1)\times \PP^1 \\
 & X=\PP^1\times \PP^1 \times \PP^1 \incl[ur]&  \\
}\]

\caption{Stratification of the projectivized Hilbert space of three qubits by secant and tangential varieties.}\label{222onion}
\end{figure}
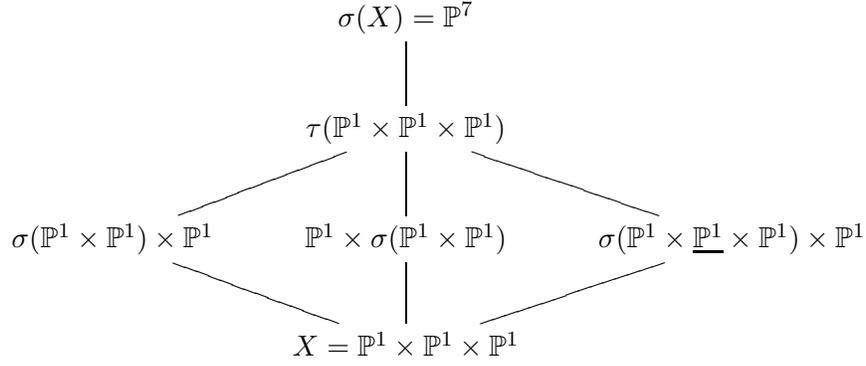

In Figure \ref{222onion} the lines represent inclusions as algebraic varieties and $\sigma(\PP^1\times\underline{\PP^1}\times \PP^1)\times \PP^1$ is a 
notation introduced in Ref\cite{HLT} to denote 
the variety of the closure of secant lines of $X$ between points of type $[u\otimes v\otimes w]$ and $[\tilde{u}\otimes v\otimes \tilde{w}]$.

Based on the previous analysis we see, that an alternative way of saying:
\begin{center}
{\em << Three qubits have two non equivalent classes of genuine entanglement, one of type $|GHZ\rangle$ and the other of type $|W\rangle$>>}
\end{center}

would be, in geometrical terms,

\begin{center}
 {\em <<The secant variety of the set of separable states $\PP^1\times\PP^1\times\PP^1$ has the expected dimension and fills the ambient space>>.}
\end{center}

In Section \ref{tripartite} we show what this last geometric formulation tells us about other types of multipartite systems.

\section{The geometry of tripartite entanglement}\label{tripartite}
Let us now consider a semi-simple complex Lie group $G$
and an irreducible $G$-module $\mathcal{H}$, i.e. a vector space such that $G$ acts on $\mathcal{H}$ and does not contain any
nontrivial submodule. We call $\mathcal{H}$ an irreducible representation of $G$.
Taking the projective space $\PP(\mathcal{H})$ there exists a unique smooth orbit $X=G.[v]\subset \PP(\mathcal{H})$ called the 
highest weight orbit\footnote{Highest weight vectors are defined after a choice of an ordering of the roots of the Lie algebra $\mathfrak{g}=Lie(G)$ which defines for each irreducible representation a unique highest 
weight\cite{F-H}. There is a bijection between the 
highest weights (up to a choice of an ordering of the root system) and the irreducible representations of $G$.}. In the case of three qubits we have $\mathcal{H}=\CC^2\otimes\CC^2\otimes\CC^2$, 
$G=SL_2(\CC)\times SL_2(\CC)\times SL_2(\CC)$ and $v=|000\rangle$, i.e. $X=G\cdot[|000\rangle]=\PP(\{|\psi\rangle \text{ separable}\}$. Let us look at an 
other standard example.
\begin{ex}\label{grass}
We consider $G=SL_6(\CC)$ and $\mathcal{H}=\Lambda^3 \CC^6$. The vector space $\mathcal{H}$ is an irreducible representation of 
 $G$ (more generaly $\Lambda^k V$ are irreducible representations of $SL(V)$ called the fundamental representations, see Ref\cite{F-H} page 221).
 Given $(e_1,e_2,\dots,e_6)$ a basis of $\CC^6$, a highest weight vector can be chosen to be 
 $v=e_1\wedge e_2\wedge e_3$. Then in this 
 case $X=SL_6(\CC)\cdot [e_1\wedge e_2\wedge e_3]\subset \PP(\Lambda^3 \CC^6)=\PP^{19}$. The variety $X$ 
 represents the set of 3-dimensional planes 
 in $\CC^6$ also known as the Grassmannian $G(3,6)$. Given a three plane of $\CC^6$ spanned by $u, v$ and $w$ we can always find $g\in SL_6(\CC)$ such 
 that $[g\cdot (e_1\wedge e_2\wedge e_3)]=[g\cdot e_1\wedge g\cdot e_2\wedge g\cdot e_3]=[u\wedge v\wedge w]$. In terms of skew-symmetric tensors, 
 the variety $X$ is the set of rank one tensors in $\mathcal{H}$.
 Now let us recall that, in quantum information theory, $\mathcal{H}=\Lambda^3\CC^6$ is the Hilbert space describing systems made of three fermions with six single-particles states.
 Then, from the above description, it follows that for three fermions with six single-particle states, $X=G(3,6)$ is the set of separable states 
 and $G=SL_6(\CC)$ is the corresponding SLOCC group.
\end{ex}
\begin{rem}
 Let $P\subset G$ denotes the stabilizer of the highest weight $v\in \mathcal{H}$, then $X=G/P$ and $X=G/P\subset \PP(\mathcal{H})$ realizes the 
 minimal embedding of the homogenous variety $G/P$. The subgroup $P\subset G$ is called a parabolic subgroup of $G$\cite{F-H}.
\end{rem}

Based on our geometric interpretation of the classes $|GHZ\rangle$ and $|W\rangle$ in the three qubits case, 
let us ask a more general question: what are the semi-simple Lie groups $G$ and the corresponding irreducible representations $\mathcal{H}$ such that 
\begin{equation}\tau(X)\subsetneq\sigma(X)=\PP(\mathcal{H})\end{equation}
where $X$ is arising from the projectivization of the highest weight vector ?

First it should be noticed that  $\sigma(X)=\sigma(G/P)=\PP(\mathcal{H})$ and $\tau(X)$ is a hypersurface of
$\PP(\mathcal{H})$ imply the ring 
of $G$-invariant polynomials on $\mathcal{H}$ is generated by the $G$-invariant irreducible polynomial vanishing on $\tau(X)$, i.e. $\CC[\mathcal{H}]^G=\CC[F]$ where 
$F$ is the irrecudible (up to scale) homogeneous polynomial defining $\tau(X)$. Indeed the fact that $\sigma(G/P)=\PP(\mathcal{H})$ says there is
a dense orbit (because the secant variety is always the closure of the orbit $G.[u+v]$ where $(u,v)$ is a general pair of points of $X$, see Ref\cite{Z2}). Therefore there 
are either no invariants or the ring 
of invariants is generated by a single polynomial. The fact that $\tau(X)$ is a $G$-invariant hypersurface tells us we are in the second case.
The representations such that $\CC[\mathcal{H}]^G=\CC[F]$ have been classified by Kac, Popov and Vinberg\cite{KPV}. 
It can be deduced from this classification which representations satisfy $\sigma(G.[v])=\PP(\mathcal{H})=\PP^{2n+1}$ where the highest weight orbit 
$G.[v]$ has dimension $n$. This is in fact done explicitly in the book of F. Zak\cite{Z2} page 51 and 53 where the author studies in detail
homogeneous varieties of small codimension in order to understand a special class of them called the Severi varieties.
We summerize the result of Zak in Table \ref{classification} and we put in perspective the corresponding systems in quantum information theory as well as the references where those cases have been separately investigated.
\begin{table}[!h]
$\begin{array}{c|c|c|c|c}
 G & \mathcal{H} & \text{\scriptsize   Highest weight orbit} & \text{\scriptsize   QIT interpretation} & \text{\scriptsize   References} \\
 \hline
 SL_2(\CC) & Sym^3(\CC^2) & v_3(\PP^1)\subset \PP^3 & \text{\scriptsize   Three bosonic} &\text{\scriptsize   Brody, Gustavsson,  Hughston.}\cite{brody2} \\
           &              &                       &\text{\scriptsize   qubits} &\text{\scriptsize   Vrana and L\'evay}\cite{VL} \\
 \hline
  SL_2(\CC)\times SO(m) & & & &\\
   \hline 
   m=3        & \CC^2\otimes Sym^2(\CC^2)                       &   \PP^1\times v_2(\PP^1)\subset \PP^5                            &  \text{\scriptsize   1 distinguished qubit} & \text{\scriptsize   Vrana and L\'evay}\cite{VL} \\
                        &                         &                                                         & \text{\scriptsize   and 2 bosonic qubits} &\\
        \hline
        m=4         &   \CC^2\otimes\CC^2\otimes\CC^2                      &  \PP^1\times\PP^1\times\PP^1\subset \PP^7& \text{\scriptsize   3 qubits} & \text{\scriptsize   D\"ur, Vidal, Cirac}\cite{Dur}\\
\hline
        m=5     & \CC^2\otimes \Lambda^{<2>} \CC^4 & \PP^1\times LG(2,4)\subset \PP^9 & \text{\scriptsize   1 distinguished qubit} & \text{\scriptsize New} \\ 
                          &                                                           &    & \text{\scriptsize   and two fermions with} & \\
                           &                                                            &    & \text{\scriptsize   4 single-particle state} &\\
                                                                                   &   & & \text{\scriptsize   and a symplectic form condition} & \\
  \hline
  m=6         &     \CC^2\otimes \Lambda^2\CC^4                    & \PP^1\times G(2,4)\subset \PP^{11} &\text{\scriptsize    1 qubit and two fermions}  & \text{\scriptsize   Vrana and L\'evay\cite{VL}}\\
                        &                         &                                      &\text{\scriptsize   with 4 single-particle states} &\\
                       \hline
  m>6 & \CC^2\otimes \CC^{m} & \PP^1\times Q^{m-2}\subset \PP^{2m-1} & \text{\scriptsize 1 qubit and 1 isotropic (m-1)-dits}          & \text{\scriptsize New}         \\
 
                        \hline
 Sp_6(\CC) & \Lambda^{<3>}\CC^6 & LG(3,6)\subset \PP^{13} & \text{\scriptsize   Three fermions } & \text{\scriptsize New} \\
           &                       &                            & \text{\scriptsize   with six single-particle states} & \\
           &                      & &\text{\scriptsize   and a symplectic form condition} & \\
           
           \hline
  SL_6(\CC) & \Lambda^3\CC^6 & G(3,6)\subset \PP^{19} & \text{\scriptsize   Three fermions} &  \text{\scriptsize   Levay and Vrana}\cite{LV}\\
          &                   &                         & \text{\scriptsize   with six single-particle states} &   \\
          \hline
   Spin_{12} & \Delta_{12} & \SS_6\subset \PP^{31} & \text{\scriptsize   Particles in } & \text{\scriptsize   S\'arosi and L\'evay}\cite{SL}\\
              &           &                      & \text{\scriptsize   Fermionic Fock spaces}\\
              \hline
              E_7 & V_{56} & E_7/P_1\subset \PP^{55} & \text{\scriptsize   Tripartite entanglement} & \text{\scriptsize   Duff and Ferrara\cite{DF}}\\
              &          &                           & \text{\scriptsize   of seven qubits} & \\
              \hline
       \end{array}$
 \caption{Classification of smooth $G$-orbits satisfying $\tau(G.[v])\subsetneq\sigma(G.[v))=\PP(\mathcal{H})$}\label{classification}
\end{table}
The notations of Table \ref{classification} are as follow:
\begin{itemize}
 \item $Sym^n V$ and $\Lambda^n V$ denote respectively the symmetric and skew-symmetric parts  of $V^{\otimes n}$.
 \item $v_k:\PP(V)\to \PP(Sym^k V)$ is the Veronese map defined by $v_k([v])=[v\circ v\circ \dots \circ v]$ and $v_2(\PP^1)$ and 
 $v_3(\PP^1)$ are curves corresponding to the images of $\PP^1$ by $v_2$ and $v_3$ also known as the conic and the twisted cubic\cite{Ha}.
 \item $Q^{n-1}\subset \PP^n$ denotes a smooth quadric in $\PP^n$.
\item The variety $LG(k,n)\subset \PP(\Lambda^{<k>}\CC^n)$ is the so-called Lagrangian Grassmannian. Given a 
non degenerate symplectic form $\omega$ on $\CC^n$, 
$LG(k,n)$ is the variety of isotropic $k$-planes in $\CC^n$ with respect to $\omega$.
\item As already  mentioned in Example \ref{grass}, the variety $G(k,n)\subset \PP(\Lambda^k  \CC^n)$ is the Grassmannian variety of $k$-planes in $\CC^n$.
\item The vector space $\Delta_{12}$ is the standard representation of the group $Spin_{12}$, i.e. the double covering of $SO(12)$, see Ref\cite{F-H}. The variety 
$\SS_6\subset \PP(\Delta_{12})$ is the corresponding highest weight orbit, called the Spinor variety. It is the variety of pure spinors\cite{chevalley}.
\item The vector space $V_{56}$ is the standard representation of the Lie group $E_7$ and $E_7/P_1$ denotes the corresponding highest weight orbit (in terms of parabolic groups $P_1$ corresponds 
to the parabolic group defined by the root $\alpha_1$).
 \end{itemize}

Table \ref{classification} provides a classification of quantum systems featuring two and only two classes of genuine entanglement of 
types $|W\rangle$ and $|GHZ\rangle$.
Although most of these systems have been studied independently by various authors of the quantum information theory community, it is interesting to
point out here that
thanks to the work of F. Zak
now a purely geometric approach allows us 
to present them in a unique classification scheme. As we will discuss it in the Appendix \ref{app} this classification 
also corresponds to the classification of Freudenthal varieties. The 
role of Freudenthal construction in the study of those quantum systems, in particular the role of Freudenthal triple systems (FTS) has been already 
understood and used  by different authors\cite{BDDER,BDFMR,VL}. The Hilbert spaces and quantum systems of our  Table \ref{classification} obtained by geometric arguments are 
the same Hilbert spaces and SLOCC groups of Table II of Ref\cite{BDDER} built from FTS. However the FTS construction does not show that this Table provides a complete classification 
of quantum systems featuring this pecular entanglement behavior.

Let us also point out that three new types of quantum systems with entanglement classes similar to the three qubit
systems appear in this classification. Their set of separables states correspond to the following three algebraic varieties:
\begin{itemize}
 \item[i)] $X=\PP^1\times LG(2,4)\subset \PP^9$,
\item[ii)] $X=LG(3,6)\subset \PP^{13}$
\item[iii)] $X=\PP^1\times Q^{m-2}\subset \PP^{2m-1}$, $m>6$.
\end{itemize}

As mentioned in Table \ref{classification}, the first system is made of a distinguished qubit and two fermions with four single-particle states satisfying a symplectic condition and the second system corresponds 
to three fermions with six single-particle states satisfying a symplectic condition. 
The last new case corresponds to a system made of a qubit and a $m-1$-dits ($m>6$)  which satisfies an isotropic condition given by a quadratic form.

\begin{rem}\label{bosons}
 The orbit structure of the projectivized Hilbert spaces $\PP(\mathcal{H})$ with the SLOCC groups $G$ of Table \ref{classification} is fully provided by Ref\cite{LM1}. 
 In particular the authors show that, except for $G=SL_2(\CC)$ and $\mathcal{H}=Sym^3(\CC^2)$ (three bosonic qubits), there are exactly $4$ orbits. The Zariski closures of those orbits can be described as follow:
 \begin{equation}
 \underbrace{X}_{\text{Separable}}\subset \underbrace{\sigma_{+}(X)}_{\text{Biseparable}}\subset \underbrace{\tau(X)}_{|W\rangle}\subset \underbrace{\sigma(X)}_{|GHZ\rangle}=\PP(\mathcal{H})
 \end{equation}
The variety $\sigma_{+}(X)$ is the closure of points of type $|\psi\rangle+|\chi\rangle$ where $|\psi\rangle$ and $|\chi\rangle$ are two 
separable states which do not form a generic pair (see Ref\cite{LM1} for the description of the isotropic condition satisfied by this pair $(|\psi\rangle,|\chi\rangle)$).
The smooth points of $\sigma_{+}(X)$ are therefore identified with biseparable states.
This variety is irreducible except in the case of three qubits where $\sigma_{+}(\PP^1\times \PP^1\times \PP^1)$ splits in three 
irreducible components (see Figure \ref{222onion}).

For three bosonic qubits, $X=v_3(\PP^1)$, the orbit structure is sligthly different. It is only made of three orbits as there is no variety such as $\sigma_{+}(X)$.
\end{rem}

We conclude this Section with a variation of our initial problem. Instead of classifying systems with two and only two classes of genuine entanglement of type 
$|W\rangle$ and $|GHZ\rangle$, let us 
consider systems having two and only two types of genuine entanglement (but not necessarly featuring $|W\rangle$ and $|GHZ\rangle$). 
\begin{ex}
 Let $\mathcal{H}=\CC^3\otimes\CC^3$, $G=SL_3(\CC)\times SL_3(\CC)$, and $X=G.[|00\rangle]=\PP^2\times\PP^2\subset \PP^8=\PP(\mathcal{H})$, i.e. $X$ is the set of separable states of two qutrits. The variety $X$ can also be identified with the projectivization 
of the rank one $3\times 3$ matrices and $\PP(\mathcal{H})$ is the projectivization of the space of $3\times 3$ matrices. 
Under the action of SLOCC group $G$, it is well known that we have only three orbits:
\begin{equation}
 X=\PP\{\text{Matrices of rank } 1\}\subset \PP\{\text{Matrices of rank }\leq 2\}\subset \PP\{\text{Matrices of rank }\leq 3\}=\PP^8
\end{equation}
Then the variety of rank less than two matrices is the secant variety of $X$ and general points of this variety correspond to the orbit of the state $|\psi\rangle=|00\rangle+|11\rangle$. But 
in this example there is no tangential variety because of dimension condition. Indeed $dim(\sigma(X))=7<2\times 4+1$ and by Proposition \ref{dim} we have $\sigma(X)=\tau(X)$.
Thus this is an example of a multipartite system with two and only two types of non equivalent entangled states but there is no entangled class of type $|W\rangle$. 
\end{ex}

It is clear form the previous example that quantum systems with only two types of genuine entangled classes which are not considered in Table \ref{classification} should correspond to 
systems whose set of separable states $X\subset \PP(\mathcal{H})$ satisfies the following geometric conditions: 
\begin{equation}\label{condition2}
 dim(\sigma(X))<2dim(X)+1 \text{ and there is a SLOCC orbit corresponding to } \PP(\mathcal{H})\backslash \sigma(X) 
\end{equation}

In turns out that the classification of homogeneous varieties $X=G/P$ under the conditions of Eq. (\ref{condition2}) can also be 
deduced from Zak's work (See Ref\cite{Z2} page 54 and 59).
We summerize this result in Table \ref{classification2}.
\begin{table}[!h]
$\begin{array}{c|c|c|c}
 G & \mathcal{H} & \text{\scriptsize   Highest weight orbit} & \text{\scriptsize   QIT interpretation}  \\
 \hline
 SL_2(\CC) & Sym^2(\CC^3) & v_2(\PP^2)\subset \PP^5 & \text{\scriptsize   Two bosons} \\
           &              &                       &\text{\scriptsize   with 3 single-particle states}  \\
 \hline
 SL_3(\CC)\times SL_3(\CC)& \CC^3\otimes \CC^{3} & \PP^2\times \PP^2\subset \PP^{8} & \text{\scriptsize Two qutrits}                    \\
              
                        \hline
 SL_5(\CC) & \Lambda^{2}\CC^5 & G(2,5)\subset \PP^{14} & \text{\scriptsize   Two fermions }  \\
           &                       &                            & \text{\scriptsize   with five single-particle states}  \\
           \hline
  E_6 & V_{27} & E_6/P_1\subset \PP^{26} & \text{\scriptsize Bipartite entanglement of three qutrit\cite{DF2}} \\
          &                   &                         &  \\
          \hline
   SL_3(\CC)\times SL_4(\CC) & \CC^3\otimes \CC^4 & \PP^2\times\PP^3\subset \PP^{11} & \text{\scriptsize  One qutrit and one 4-qudit } \\
              \hline
            SL_7(\CC) & \Lambda^2\CC^7 & G(2,7)\subset \PP^{20} & \text{\scriptsize   Two fermions} \\
              &          &                           & \text{\scriptsize  with 7 single-particle states} \\
              \hline
       \end{array}$
 \caption{Classification of smooth $G$-orbits with orbit structure $G.[v]\subset\sigma(G.[v])\subset\PP(\mathcal{H})$}\label{classification2}
\end{table}
The notations for Table \ref{classification2} are as follows:

\begin{itemize}
 \item $V_{27}$ is the standard representation of $E_6$ and $E_6/P_1$ is the highest weight orbit.
\end{itemize}

The first four varieties of Table \ref{classification2} are the so-called Severi varieties studied by F. Zak\cite{Z2}. In terms of entanglement Tables \ref{classification} and \ref{classification2}
lead to the following result.
\begin{theorem}
 The pure quantum systems having two and only two type of genuine entanglement classes are classified by Tables \ref{classification} and \ref{classification2}.
\end{theorem}

\begin{rem}
 It should be pointed it out that the composite quantum systems of Table \ref{classification} are all tripartite systems 
 (except in the case of $\mathcal{H}=\CC^2\otimes \CC^m$  with $G=SL_2(\CC)\times SO(m)$ for $m>6$), while the composite systems of Table \ref{classification2} are all bipartite systems.
 This will be emphasized in Appendix \ref{appendix} when we refer to a uniform geometric parametrization of the varieties of separable states given by Ref\cite{LM1}.
\end{rem}

 

 \section{Conclusion}
 By means of algebraic geometry in this paper we intended to provide a uniform description of pure quantum systems featuring a classification of entanglement 
 types similar to 
 the famous case of three qubits. More precisely we explained how a geometric interpretation of what the $|W\rangle$ and $|GHZ\rangle$ states are, allows us to 
 use  results of algebraic geometry and invariant theory to give an explicit list (Table \ref{classification}) of all Hilbert spaces, with 
 the corresponding SLOCC group, such that the only types of genuine 
 entangled states are the exact analogues of the $|W\rangle$ and $|GHZ\rangle$ states. It turns our that this list of separable states for those Hilbert spaces correponds to the list of subexceptional
 Freudenthal varieties. 
 Those varieties
 have a strong connexion with exceptional simple Lie algebras (as fundamental subadjoint varieties). They also admit a uniform description as image of the same 
 rational map (Pl\"ucker embedding) over 
 different composition algebras. This map found by Ref\cite{LM1} is described in Appendix \ref{appendix}. The translation of the work of algebraic geometers\cite{LM1,LM2,LM3,Z2} we manage to do to quantum information theory language
 could be summerize in the following sentence: <<{\em Three fermions with $6$ single-particle states over composition algebras can 
 be entangled in two different ways}>>. This sentence includes all known cases of tripartite systems having a similar orbit structure as the 
 three qubit case.

 \appendix
 
\section{The tripartite entanglement and the Freudenthal varieties}\label{app}
The algebraic varieties of Table \ref{classification} have been studied in the mathematics literature as 
 the fundamental 
subadjoint varieties or the Freudenthal varieties. In the early 2000, Landsberg and Manivel investigated 
the geometry of those varieties in a series of papers\cite{LM1,LM2,LM3}. Their goal was to establish new connections between representation theory and algebraic geometry.
In this Appendix we collect some results and descriptions of this sequence of varieties which we believe to be  relevant for quantum information theory.
\subsection{The subadjoint varieties}
Let us consider $\mathfrak{g}$ a complex simple Lie algebra of type $B_n,D_n, G_2, F_4, E_6, E_7$ and $E_8$ (i.e. all complex simple Lie algebras except those of type $A_n$ and $C_n$).
They correspond to the fundamental simple Lie algebras, i.e. Lie algebras whose adjoint representation is fundamental\cite{LM1}.
Then let $X_G\subset \PP(\mathfrak{g})$ be the highest weight orbit for the adjoint representation of the corresponding Lie group  $G$. Consider $\tilde{T}_x X_{G}$ the embedded tangent space at any point $[x]$ of the 
homogeneous variety $X_G$. Then $Y=X_G\cap T_{x} X_{G}$ is a homogenous variety. Table \ref{subadjoint} gives the correspondence between the Lie algebras $\mathfrak{g}$ and the homogenenous varieties $Y$.
\begin{table}[!h]
\[\begin{array}{c|c}
   Y\subset \PP^{n} & \mathfrak{g} \\
\hline
 v_3(\PP^1)\subset \PP^3 & \mathfrak{g}_2\\
\PP^1\times \QQ^{m-4} \subset \PP^{2m-5} & \mathfrak{so}_m\\
LG(3,6)\subset \PP^{13} & \mathfrak{f}_4\\
G(3,6)\subset \PP^{19} & \mathfrak{e}_6\\
\SS_6\subset \PP^{31} & \mathfrak{e}_7\\
E_7/P_1\subset \PP^{55} & \mathfrak{e}_8
  \end{array}\]
\caption{Fundamental subadjoint varieties}\label{subadjoint}
\end{table}
The sequence of algebraic varieties corresponding to quantum multipartite systems featuring only the two types of genuine entanglement $|W\rangle$ and 
$|GHZ\rangle$ are connected to fundamental adjoint representations of Lie algebras by this construction. 
Moreover in Ref\cite{LM2} Landsberg and Manivel prove the existence of a rational map 
of degree $4$ which allows from
the knowledge of $Y$ to reconstruct the adjoint varieties $X_G\subset \PP(\mathfrak{g})$ and thus  to recover
the structure of the Lie algebra $\mathfrak{g}$. 

To illustrate the construction of this rational map, let us detail one  example.

\begin{ex}\label{rationnalmap}
 Let $Y=G(3,6)\subset \PP^{19}=\PP(V)$ be the variety of separable states for a system made of three fermions with six single-particule states. Let us denote by 
 $\{x_1,\dots, x_{20}\}$ a dual basis of $V$. 
 Then embedded linearly $\PP(V)=\PP^{19}\subset_{\{x_0=0\}} \PP^{20}\subset_{\{x_{21}=0\}} \PP^{21}$ and consider the rational map
 $\phi: \PP^{21}  \to \PP^{77}$ defined by \[\phi([x_0,\dots,x_{21}])=[x_0^4,x_0^3x_{21}, x_0^3 x_i,x_0^2 I_2(Y), x_0x_{21}x_i-x_0 I_3(\tau(Y)_{sing}),x_0^2x_{21}^2-I_4(\tau(Y))]\]
 
 where $1\leq i\leq 20$, $I_k(Z)$ denotes a set of generators of the ideal of degree $k$ polynomials defining $Z$ and $\tau(Y)_{sing}$ is the subvariety of singular points of $\tau(Y)$. 
 Then $\phi(G(3,6))=X_{E_6}$, i.e. $\phi$ maps the set of separable three fermions with six single-particules states to the $E_6$ adjoint variety.
 The $E_6$ adjoint variety contains the information defining the Lie algebra $\mathfrak{e}_6$ as we have $<X_{E_6}>=\PP(\mathfrak{e}_6)$, i.e. the linear span fills the full space and the 
 algebraic structure can be recovered\cite{LM2} from the geometry of $X_{E_6}$.

\end{ex}

One sees from the previous example that the Lie algebra $\mathfrak{e}_6$ can be reconstructed from the defining equation of $\tau(G(3,6))$, i.e. the unique (up to a multiplication by a scalar) 
SLOCC invariant irreducible quartic on $\mathcal{H}=\Lambda^3(\CC^6)$. Indeed the ideal of degree three polynomials vanishing on the singular locus 
of $\tau(G(3,6))$ is generated by the derivatives of the quartic invariant and the ideal of degree two polynomials defining $G(3,6)$ is spaned 
by the second derivative of the quartic invariant.
But this quartic invariant is known in the context of entanglement as the analogue for three bosons of the $3$-tangle\cite{VL}.

Therefore in the context of entanglement Landsberg and Manivel's construction tells us that Table \ref{subadjoint} can be read as follows: Consider a fundamental 
Lie algebra $\mathfrak{g}$ and the corresponding mutlipartite quantum system $Y$. Then $\mathfrak{g}$ can be reconstructed from the knowlegde of the unique irreducible 
SLOCC invariant of degree $4$ (i.e. the generalization of the  $3$-tangle). This is another approach to construct Lie algebra from qubits\cite{CB}.

\subsection{The Freudenthal subexceptional series}\label{appendix}
The subadjoint varieties also appeared in the work of Landsberg and Manivel in their geometric investigation of the so-called Freudenthal
magic square. Let us recall that the Freudenthal magic square is a square of semi-simple Lie algebras due to Freudenthal\cite{Freu} and Tits\cite{T} 
 obtained from a pair
of composition algebras 
$(\AA,\BB)$ (where $\AA$ and $\BB$ are the complexification of $\RR, \CC, \HH$, the quaternions or $\OO$, the octonions) by the following construction:
\begin{equation}
 \mathfrak{g}=Der(\AA)\oplus(\AA_0\otimes J_3(\BB)_0)\oplus Der(J_3(\BB))
\end{equation}

where $\AA_0$ denotes the space of  imaginary elements, 
$J_3(\BB)$ denotes the Jordan algebra of $3\times 3$ Hermitian matrices over $\BB$ and $J_3(\BB)_0$ is the subspace of traceless matrices of $J_3(\BB)$. For an algebra $A$, 
$Der(A)$ is the derivation
of $A$, i.e. the Lie algebra of the automorphism group of $A$.

The Freudenthal magic square is thus given by:
\begin{table}[!h]
\begin{center}
\begin{tabular}{c|cccc}
  & $\RR$ & $\CC$ & $\HH$ & $\OO$\\
  \hline
  $\RR$ & $\mathfrak{s}\mathfrak{o}_3$ &$\mathfrak{s}\mathfrak{l}_3$ & $\mathfrak{s}\mathfrak{p}_6$  & $\mathfrak{f}_4$ \\
  $\CC$ & $\mathfrak{s}\mathfrak{l}_3$ &$\mathfrak{s}\mathfrak{l}_3\times \mathfrak{s}\mathfrak{l}_3$ & $\mathfrak{s}\mathfrak{l}_6$  & $\mathfrak{e}_6$ \\
  $\HH$ &  $\mathfrak{s}\mathfrak{p}_6$ &$\mathfrak{s}\mathfrak{l}_6$ & $\mathfrak{s}\mathfrak{o}_{12}$  & $\mathfrak{e}_7$ \\
  $\OO$ & $\mathfrak{f}_4$ & $\mathfrak{e}_6$ & $\mathfrak{e}_7$ & $\mathfrak{e}_8$ \\
  \hline
\end{tabular}
\end{center}
\caption{The Freudenthal magic square}
\end{table}

The relevence of Freudenthal construction to the study of entanglement has been pointed out by various authors\cite{BDDER,VL}. However
the geometric contribution has not been completely explained so far in the context of quantum information theory. The geometric version of the Freudenthal magic square given in Ref\cite{LM2,LM3} is the following 
square of 
homogeneous varieties:

\begin{table}[!h]
\begin{center}
\begin{tabular}{c|cccc}
  & $\RR$ & $\CC$ & $\HH$ & $\OO$ \\
  \hline
  $\RR$ & $v_2(Q^1)$ & $\PP(T(\PP^2))$ & $ LG(2,6)$  & $E_6/P_1\cap H$ \\
  $\CC$ & $v_2(\PP^2)$ &$\PP^2\times\PP^2$ & $G(2,6)$  & $E_6/P_1$ \\
  $\HH$ &  $LG(3,6)$ &$G(3,6)$ & $\SS_6$  & $ E_7/P_7$ \\
  $\OO$ & $X_{F_4}$ & $X_{E_6}$ & $ X_{E_7}$ & $X_{E_8}$ \\
  \hline
\end{tabular}
\end{center}
\caption{The Geometric magic square}
\end{table}

The geometric magic square has the property that each homogeneous variety of the square is homogenous for the corresponding Lie group in the Freudenthal magic square. Moreover each variety of 
a given row can be recovered as a section (tangential or linear) of the next one.
The connexion with composition algebras leads 
Landsberg and Manivel to formulate a geometrical uniform description of the varieities of the third row (the one relevant for the classification of Table \ref{classification}) as Grassmannians over the composition algebras. 

It is well known that the variety $G(3,6)$ of Example \ref{rationnalmap} can be parametrized by the so-called Pl\"ucker map\cite{Ha}. Let $v_1,v_2$ and $v_3$ be three complex vectors 
 defining a three plane in $\CC^6$.
The coordinates can be chosen so that $v_1=[1:0:0:0:0:0]$, $v_2=[0:1:0:0:0:0:0]$ and $v_3=[0:0:1:0:0:0]$. 
 Let $[\tilde{v}_1\wedge \tilde{v}_2\wedge \tilde{v}_3]$ be a three plane in the neighborhood of $[v_1\wedge v_2\wedge v_3]$. One can choose $\tilde{v}_1=[1:0:0:a_{11}:a_{12} :a_{13}]$, 
 $\tilde{v}_2=[0:1:0:a_{21}:a_{22} :a_{23}]$ and $\tilde{v}_3=[0:0:1:a_{31}:a_{32} :a_{33}]$. Locally the variety $G(3,6)$
is parametrized in the neighborhood of $[v_1\wedge v_2\wedge v_3]$ by 
\begin{equation}\label{plucker}
 \phi(1,P)=(1,P,com(P),det(P))
\end{equation}

where $P$ is the matrix $P=(a_{ij})$ and $com(P)$ is its comatrix. The map $\phi$ is the Pl\"ucker map.

An alternative description of $G(3,6)$ can be given by considering $P\in J_3(\AA)$ where $\AA=\CC\oplus \CC$ is the complexification of $\CC$, i.e. 
$P=\begin{pmatrix}
    \alpha & x_1 & x_2\\
    \overline{x_1} & \beta & x_3\\
    \overline{x_2} & \overline{x_3} & \gamma
   \end{pmatrix}$ with $\alpha, \beta, \gamma\in \CC$ and $x_1, x_2, x_3\in \CC\oplus \CC$.
Then to recover the same
parametrization one needs to require that the three row vectors defining the matrix $(I_3|P)$ are orthogonal with respect 
to the symplectic form $\omega=\begin{pmatrix}
                                                                                                                                                 0 & I_3\\
                                                                                                                                                 -I_3 & 0
                                                                                                                                                \end{pmatrix}$, i.e. the corresponding $3$-plane is isotropic for $\omega$.
                                                                                                                                                
Under the symplectic condition, one has $G(3,6)=LG_{\CC\oplus \CC}(3,6)$.
                                                                                                                                                
Similarly the Pl\"ucker map of Eq. (\ref{plucker}) can be defined for $P\in J_3(\AA)$, with $\AA$ one of the three other complex composition algebras. 
Then if we denote by $\AA=\CC$, $M_2(\CC), \OO_\CC$ the complexifications of $\RR, \HH, \OO$, 
Landsberg and Manivel proved\cite{LM2} that
the varieties of the third row can all be interpreted as $LG_{\AA}(3,6)$, i.e.
\begin{equation}
\left.\begin{array}{c}
 LG(3,6)=LG_\CC(3,6)\\
  G(3,6)=LG_{\CC\oplus \CC}(3,6)\\
  \SS_{6}=LG_{M_2(\CC)}(3,6) \\
  E_7/P_7=LG_{\OO_{\CC}}(3,6)  
\end{array}\right\}\begin{array}{c}
\text{Three $\AA$-fermions with six single-particle}\\
\text{states satisfying a symplectic condition}
\end{array}
\end{equation}

 Moreover if we consider the case  $P\in J_3(\underline{-1})=\{\begin{pmatrix}
                                                                                                                 \alpha & 0 & 0\\
                                                                                                                 0 & \alpha & 0\\
                                                                                                                 0 & 0& \alpha
                                                                                                                \end{pmatrix}, \alpha\in \C\}$ (notations of Ref\cite{LM2}) and 
                                                                                                                the case $P\in J_3(\underline{0})=\{\begin{pmatrix}
                                                                                                                                                                                                                              \alpha & 0 & 0\\
                                                                                                                                                                                                                              0 & \beta & 0\\
                                                                                                                                                                                                                              0 & 0 & \gamma
                                                                                                                                                                                                                             \end{pmatrix}, \alpha, \beta,\gamma\in \CC\}$,
one obtains a similar Pl\"ucker parametrization of $v_3(\PP^1)=LG_{\underline{-1}}(3,6)$ and $\PP^1\times \PP^1\times \PP^1=LG_{\underline{0}}(3,6)$.

An important consequence for quantum information theory is that this geometric interpretation of the varieties of the extended third row as Lagrangian Grassmannians over $\AA$ say that
all quantum systems, which feature only the states $|W\rangle$ and $|GHZ\rangle$ as their  genuine entangled classes, are 
{\em tripartite systems of indistingushable particles with six-single particle states 
with coefficients  in a complex composition algebra  satisfying a symplectic condition}.

\begin{rem}
 Similarly a description of the first four varieties of separable states of Table \ref{classification2} as Lagrangian Grassmannians $LG(\AA^2,\AA^6)$ is given in Ref\cite{LM2}. 
 Those varieties correspond to the second row of the geometric magic square.
\end{rem}


\begin{thebibliography}{99}
\bibitem{bell} Bell J. S. ``On the Einstein-Podolsky-Rosen paradox.'' Physics 1, no. 3 (1964): 195-200.
\bibitem{BDD}  Borsten L.,  Dahanayake D.,  Duff M. J., Marrani A. and Rubens W., ``Four-Qubit Entanglement Classification from String Theory'', Phys. Rev. Lett. {\bf 105}, 100507 (2010). 
\bibitem{BDDER}  Borsten L.,  Dahanayake D.,  Duff M. J., Ebrahim H. and Rubens W., ``Freudenthal triple classification of three-qubit entanglement'', Physical Review A 80, no 3 (2009):032326.
	
\bibitem{BDDER2} Borsten L., Dahanayake D., Duff M. J., Ebrahim H., and Rubens W., ``Black holes, qubits and octonions'', Physics Reports 471, no. 3 (2009): 113-219.
\bibitem{BDFMR} Borsten L., Duff M. J., Ferrara S., Marrani A. and Rubens, W., ``Explicit orbit classification of reducible Jordan algebras and Freudenthal triple systems''. arXiv preprint arXiv:1108.0908. (2011).

 \bibitem{brody2}  Brody D.C.,  Gustavsson A.C.T. and  Hughston L.P., ``Entanglement of three-qubit geometry'', J. Phys. Conf. Ser. {\bf 67}, 012044 (2007).
\bibitem{BDL} Borsten L., Duff M. J., and Levay P., ``The black-hole/qubit correspondence: an up-to-date review.'', arXiv preprint arXiv:1206.3166 (2012).
%
%
%
%
\bibitem{chevalley} Chevalley C., {\em The Algebraic Theory of Spinors and Clifford Algebras: Collected Works}. Vol. 2. Springer, 1997.

\bibitem{CB} Cerchiai B. L. and van Geemen B. ``From qubits to E7'', Journal of mathematical physics, 51(12), 122203 (2010).
%


	

\bibitem{CKW} Coffman V., Kundu J. and Wootters W. K., ``Distributed entanglement'', Phys. Rev. A61, 052306 (2000).
\bibitem{DF} Duff M. J., and  Ferrara S. ``E 7 and the tripartite entanglement of seven qubits'', Physical Review D 76.2 (2007): 025018.
\bibitem{DF2} Duff M. J., and Ferrara, S.  ``E 6 and the bipartite entanglement of three qutrits'', Physical Review D, 76 (12), (2007): 124023.
\bibitem{Dur}  D\"ur W.,  Vidal G. and  Cirac J. I., ``Three qubits can be entangled in two inequivalent ways'', Phys. Rev. A {\bf  62}, 062314 (2000).
%

\bibitem{FHa} W. Fulton, J. Hansen, ``A connectedness theorem for projective varieties, with applications to intersections and singularities of mapping'', Ann. Math. {\bf 110} (1979), 159-166.
 \bibitem{F-H} W. Fulton, J. Harris, {\em Representation Theory},
     Graduate Text in Mathematics, Springer 1991.
\bibitem{Freu} H. Freudenthal, ``Beziehungen der $e_7$ und $\e _8$ zur Oktavenebene, I, II'', Indag. Math. 16 (1954), 218-230, 363-368. III, IV, Indag. Math. 17 (1955), 151-157, 277-285. V - IX, Indag. Math. 21 (1959), 165-201, 447-474. X, XI, Indag. Math. 25 (1963) 457-487. 

%
%
 \bibitem{Ha} J. Harris, {\em Algebraic Geometry: a first course}, Graduate Texts in Mathematics {\bf{133}} Springer 1992.
%
%
%
%
%

\bibitem{HLT} Holweck F., Luque J.-G., Thibon J.-Y., ``Geometric descriptions of entangled states by auxiliary varieties'', Journal of Mathematical Physics {\bf 53}, 102203 (2012).
\bibitem{HLT2} Holweck F., Luque J.-G., Thibon J.-Y., ``Entanglement of four qubit systems: a geometric atlas with polynomial compass I (the finite world)'', Journal of Mathematicla Physics {\bf 55} (1), 012202 (2014). 


 \bibitem{Lan} T. Ivey, J. M. Landsberg, {\em  Cartan for beginners: Differential Geometry via Moving Frames and Exterior Differential Systems}, Graduate Studies in Mathematics {\bf{61}} 2003.
%


%

 \bibitem{KPV} Kac V. G., Popov V. L. and Vinberg E. B., ``Sur les groupes alg\'ebriques dont l'alg\`ebre des invariants est libre'', C.R. Acad. Sci. Paris, ser A. {\bf 283} (1976), 875-878.
 
 \bibitem{KL}  Kallosh R. and Linde A., ``Strings Black Holes and Quantum Information'',
Phys. Rev. D73 104033 (2006).

\bibitem{KSV} Kitaev A. Y., Shen A. and  Vyalyi M. N., Classical and quantum computation. No. 47. American Mathematical Soc., 2002.

%
%
%
%
%
%
%

 \bibitem{Lan3} Landsberg J. M., {\it Tensors: Geometry and applications}, Vol. 128. Amer Mathematical Society, 2011.
%
%

\bibitem{LM1} Landsberg J. M. and Manivel L., ``The projective geometry of Freudenthal's magic square'', Journal of Algebra 239.2 (2001): 477-512.

\bibitem{LM2} Landsberg J. M. and Manivel L., ``Construction and classification of complex simple Lie algebras via projective geometry'', Selecta Mathematica 8.1 (2002): 137-159.

\bibitem{LM3} Landsberg, J. M., and Manivel L. ``Representation Theory and Projective Geometry'', in Algebraic Transformation Groups and Algebraic Varieties, Encyclopedia of Mathematical Sciences. Subseries Invariant Theory and Algebraic Transformation Groups, Vol. III,
Springer-Verlag, 2004.

 \bibitem{LePai}Le Paige C., ``Sur la th\'eorie des formes binaires \`a plusieurs s\'eeries de
 variables'', Bull. Acad. Roy. Sci. Belgique  {\bf 2} (3), 40-53 (1881).

 	
\bibitem{Levay1} L\'evay P. ``Three-qubit interpretation of BPS and non-BPS STU black holes.'', Physical Review D 76.10 (2007): 106011.

\bibitem{LV} L\'evay P., and Vrana P. ``Three fermions with six single-particle states can be entangled in two inequivalent ways''. Physical Review A, 78(2), 022329. (2008)
%
%
%
%
%
\bibitem{My} Miyake A., ``Classification of multipartite entangled states by multidimensional determinants'', Phys. Rev. A {\bf 67}, 012108 (2003).
%
 \bibitem{My3} Miyake  A., ``Multipartite Entanglement under Stochastic Local Operations and Classical Communication'', Int. J. Quant. Info.  {\bf 2}, 65-77 (2004).
%
 \bibitem{My2} Miyake A. and Verstraete F., ``Multipartite entanglement in $2\times 2\times n$ quatum systems'', Phys. Rev. A {\bf 69}, 012101 (2004).
\bibitem{nielsen} Nielsen M. A. and Chuang I. L., {\em Quantum computation and quantum information}. Cambridge university press, 2010.

%

\bibitem{PV} Popov V., Vinberg E., ``Invariant theory'', Algebraic geometry IV. Springer Berlin Heidelberg, 1994. 123-278.
%

\bibitem{SL} S\'arosi  G., and L\'evay P., ``Entanglement in fermionic Fock space'', Journal of Physics A: Mathematical and Theoretical 47.11 (2014): 115304.
\bibitem{SL2} S\'arosi G., and L\'evay P., ``Entanglement classification of three fermions with up to nine single-particle states'', Physical Review A 89, no. 4 (2014): 042310.



%
%

\bibitem{T} Tits. J. ``Alg\`ebres alternatives, alg\`ebres de Jordan et alg\`ebres de Lie exceptionnelles I : Construction.'' Indag. Math. {\bf 28} 223-237 (1966)

%
%
%

\bibitem{VL} Vrana P., and  L\'evay P., ``Special entangled quantum systems and the Freudenthal construction'', Journal of Physics A: Mathematical and Theoretical 42.28 (2009): 285303.

\bibitem{VDMV}  Verstraete F.,  Dehaene F.,  De Moor B. and  Verschelde H., ``Four qubits can be entangled in nine different ways'', Phys. Rev.. A {\bf 65}, 052112 (2002).



%
%

 \bibitem{Z2} F. Zak, {{\em Tangents and Secants of Algebraic Varieties}, AMS Translations of mathematical monographs {\bf{127}} 1993.}

 \end{thebibliography}
\end{document}